# FUTURE COLLIDERS BASED ON A MODULATED PROTON BUNCH DRIVEN PLASMA WAKEFIELD ACCELERATION

Guoxing Xia[1,2], Allen Caldwell[3], Patric Muggli[3]
1. School of Physics and Astronomy, the University of Manchester, Manchester, U.K.
2. The Cockcroft Institute, Daresbury, U.K.
3. Max Planck Institute for Physics, Munich, Germany

*Abstract*

Recent simulation shows that a self-modulated high energy proton bunch can excite a large amplitude plasma wakefield and accelerate an externally injected electron bunch to the energy frontier in a single stage acceleration through a long plasma channel. Based on this scheme, future colliders, either an electron-positron linear collider ($e^+$-$e^-$ collider) or an electron-hadron collider ($e$-$p$ collider) can be conceived. In this paper, we discuss some key design issues for an $e^+$-$e^-$ collider and a high energy $e$-$p$ collider, based on the existing infrastructure of the CERN accelerator complex.

## INTRODUCTION

It has been recently proposed to use a high energy proton bunch as the drive beam to excite a plasma wakefield and accelerate the electron beam to high energies [1,2]. The advantages of using the proton beam as driver, compared to other drive pulses like electron bunch and laser pulse lie in the fact of the availability of the high energy proton beam and of the extremely high energies stored in current proton beams. For instance, the energy stored at a 1 TeV LHC-like proton bunch is in generally more than two orders of magnitude higher than that of the nowadays' electron bunch and the laser pulse. The particle-in-cell (PIC) simulation shows that for a 1 TeV LHC-like proton bunch, if compressed longitudinally to 100 microns in length, may act as a good driver and can excite a plasma wave with a field amplitude of ~2 GeV/m. Surfing on the right phase, a bunch of electrons can gain energies up to 600 GeV in a single passage of a 500 m long plasma [1]. If this scheme can be demonstrated experimentally, it will point us to a way for the TeV colliders design based at existing TeV proton machines, e.g. the CERN accelerator complex.

However, one hurdle in the above scheme is the proton bunch compression. Bunch compression via magnetic chicane is one of widely used methods to compress the electron bunch to sub-millimetre scale. If we adopt this idea and intend to compress the proton bunch to sub-millimetre, while still keeping the bunch charge as constant is not a trivial thing. It turns out that a large amount of RF power is needed to provide the energy chirp along the bunch and the large dipole magnets are required to offer the energy-path correlation beam pass. Simulation shows that the 4 km RF cavities are required to do this task [4]. This seems not practical. And then, do we have other options to compress the bunch? Yes, ask plasma for help.

It has long been known that a long laser pulse can be modulated by a high density plasma. This so called self-modulated laser wakefield acceleration (SM-LWFA) has demonstrated the large wakefield amplitude of 100 GeV/m [5]. In this scenario, the SM process occurs due to forward Raman scattering, i.e., the laser light scatters on the noise at the plasma period, that results in a wave downshifted by the plasma frequency. The two waves then beat together to drive the plasma wave. Eventually the long pulse is split into many ultra-short slices with length of each half of plasma wavelength and with each separated by a plasma wavelength (note that the plasma wavelength is inversely proportional to the square root of the plasma density). Similarly, when a long proton bunch shoots into a plasma, it can also be modulated by the wakefield it produces. It takes some time for the modulation to occur, however, once the modulation is fully set up, those proton bunch slices will drive wakefield resonantly and the field will add up coherently. Recent simulation shows that the maximum wakefield amplitude from a modulated proton bunch is comparable to that of a short bunch driver. For example, an LHC beam with a beam energy of 7 TeV, a bunch intensity of $1.15 \times 10^{11}$ and an rms bunch length of 7.55 cm can excite a wakefield with maximum amplitude of ~1.5 GeV/m by working in self-modulation regime. An externally injected electron bunch can be accelerated up to 6 TeV after propagating through a 10 km uniform plasma [3]. This indicates that we may achieve a very high energy electron beam by using today's long and high energy proton bunch as driver (given that we could make such long plasma cell for the experiment). In this scenario, we can start thinking of the collider design based at the existing high energy proton machines.

In this paper, we discuss some key issues in designing the multi-TeV colliders in which an electron positron linear collider and a high energy electron hadron collider based on a modulated proton driven plasma wakefield acceleration scheme are taken into account. These colliders will complement the current LHC and can be fully accommodated at the current sites of proton machines, e.g. CERN site.

## AN ELECTRON POSITRON COLLIDER

As we mentioned earlier, a modulated high energy proton bunch can produce a high amplitude plasma wakefield and accelerate a trailing electron bunch to the energy frontier in a single stage of acceleration. The latest simulation shows that a positron beam can also be accelerated in the wakefield from a modulated proton bunch [6]. We can therefore conceive of a TeV $e^+$-$e^-$ collider design based on this self-modulation scheme. Simulation indicates that in this case the excited wakefield always shows a decay pattern. This is mainly due to the phase shift between the beam slices and the phase of the produced wakefield. To overcome the field decay, a plasma density step-up procedure is introduced to compensate the phase change and eventually a stable and nearly constant field is achieved. Recent study shows that in this case the acceleration process is almost linear [3]. If we could make a 2 km plasma (take into account the LHC radius of 4.3 km and the focusing the beam before the plasma and the beam deliveries and IPs may also need some space), we may achieve the 1 TeV electron and positron beams from the LHC beams. Figure 1 shows a schematic layout of a 2 TeV centre-of-mass energy $e^+$-$e^-$ collider located at the LHC accelerator complex (the plasma accelerators is marked in red).

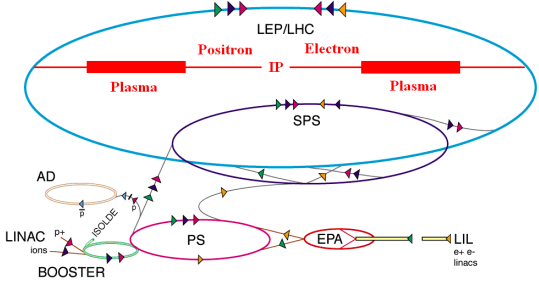

Figure 1: Schematic layout of a multi-TeV electron-positron linear collider based on a modulated proton-driven plasma wakefield acceleration.

In this scheme, the plasma channel for electron and positron is 2 km, respectively. Before entering the plasma cell, a focusing channel is designed to focus the proton beam so as to match the plasma focusing force. Then the proton bunch shoots into the plasma and excites the wakefields. We expect that after a few metres propagation in the plasma and together with a plasma density step-ups, a full modulation is finally set up and a constant wakefield is excited. An externally injected electrons and positrons will be injected into the plasma with a right phase (e.g. by tuning the position and angle of the injected beams, etc.) and sample the wakefield and are accelerated. After the 2 km plasma, a 1 TeV electron beam and positron beam will be produced (we assume that the average accelerating field in the plasma is 0.5 GeV/m, which is quite modest according to simulation results given in [3]). A 2 km beam delivery system for both electrons and positrons will transport and focus the electrons and positrons to the interaction points (IP) for collisions. After interactions with the plasmas, the proton bunches will be extracted and dumped. These spent protons may also be recycled by some cutting edge technologies for reuse or used to trigger the nuclear power plants.

## AN ELECTRON PROTON COLLIDER

It has long been known that lepton-hadron collisions have been playing an important role in exploration of deep insides of matter. For example, the quark-parton model originated from investigation of electron-nucleon scattering. To build a collider based on a moduated proton driven plasma wakefield acceleration, we could also think about an electron-hadron collider based at the CERN accelerator complex. The advantage of this design is based on the fact that the plasma-based option may be more compact and cheaper since it does not need to build a new electron accelerator.

In one of our designs, the SPS beam is used as the drive beam for plasma wakefield excitation. The reason for that is due to the long LHC beam ramping time (20 minutes). During the LHC beam energy ramping up from 450 GeV to 7 TeV, the SPS can prepare the drive beams (ramping time of LHC preinjectors is about 20 seconds) for the plasma and then excite the wakefield and accelerate the electron beam. When the accelerated electron beam is ready, it can be delivered to the collision points in the LHC tunnel for electron-proton collision. PIC simulation shows that working in the self-modulation regime, the wakefield amplitude of 1 GeV/m can be achieved by using the SPS beam at an optimum condition (both the beam and plasma parameters are optimized) [7]. Firstly the SPS beam needs to be guided to the plasma cell. Prior to the plasma cell, a focusing channel is designed to match the beam with the plasma beta function. A 170 m long plasma cell is used to accelerate the electron beam energy to 100 GeV. The energetic electrons are then extracted to collide with the circulating 7 TeV proton beam.

The centre-of-mass energy in this case is given by
$$\sqrt{s} = 2\sqrt{E_e E_p} = 1.67 \text{ TeV}$$
where $E_e$ and $E_p$ are the energy of electrons and protons respectively. The centre of mass energy in this design is about a factor of 1.2 higher than the current LHeC design (electron beam energy of 60 GeV in the LHeC design [8]) and a factor of 5.5 higher than the late HERA [9].

The luminosity of a linac-ring type $e$-$p$ collider for round and transversely matched beam is given by [10]
$$L_{ep} = \frac{1}{4\pi} \frac{P_e}{E_e} \frac{N_p}{\varepsilon_p^N} \frac{\gamma_p}{\beta_p^*}$$
where $P_e$ is electron beam power, $E_e$ is electron beam energy, $N_p$ is the number of particles in proton bunch, $\varepsilon_p^N$ is the normalized emittance of proton beam, $\gamma_p$ is the Lorentz factor and $\beta_p^*$ is beta function of proton beam at interaction point. The electron beam power is given by
$$P_e = N_e E_e n_b f_{rep}$$
where $N_e$ is the number of particles in electron bunch, $n_b$ is the number of bunches in linac pulse and $f_{rep}$ is the repetition rate of the linac. Using the LHC beam parameters, for example, $N_p$=1.15×10$^{11}$, $\gamma_p$=7460,

$\beta_p^* = 0.1$ m, $\varepsilon_p^N = 3.5$ μm and assuming the electron beam parameters as follows: $N_e=1.15\times10^{10}$ (10% of the drive beam charge), $E_e=100$ GeV, $n_b = 288$ and $f_{rep} \approx 15$, the calculated luminosity of the electron proton collider is about $1\times 10^{30}$cm$^{-2}$s$^{-1}$ for this design. However, if one can increase the electron bunch intensity and the repetition rate, it may be possible to get a higher luminosity electron proton collider.

## PHASE SLIPPAGE

When electron gains energy from the wakefield, it can reach the relativistic energy regime very quickly. As we know, the relativistic gamma factor for a 1 TeV proton is smaller than that for an electron with energy of only 1 GeV. Therefore when the proton drive beam loses its energy, its velocity may be smaller than the velocity of electron. For the collider design based on proton-driven plasma wakefield acceleration, the phase slippage may become significant when a very long acceleration channel is used.

The relative path difference due to the velocity difference is given by

$$\Delta L = \frac{1}{c}\int_0^L |v_i(s) - v_e(s)| ds$$

where $c$ is the speed of light, $L$ is the length of acceleration channel. The subscripts $i$ and $e$ denote the driving proton and witness electron bunch respectively.

Using the phase slippage formula given in [11], we have

$$\delta = k_p \frac{\Delta L}{L} \approx \frac{1}{eE_{acc}/m_e c\omega_p}(\gamma_e - \gamma_{e0})\left[1 - \frac{(\gamma_i - \gamma_{i0})}{\left(\sqrt{\gamma_i^2-1} - \sqrt{\gamma_{i0}^2-1}\right)}\right] \quad (1)$$

where $\gamma_e, \gamma_{e0}, \gamma_i, \gamma_{i0}$ denote the gamma factors for the final electron beam, initial electron beam, final proton beam and initial proton beam, respectively. $k_p = \omega_p/c$ is the plasma wave number, $\omega_p = (n_p e^2/\varepsilon_0 m_e)^{1/2}$ is the plasma electron frequency, $n_p$, $e$, $c$, $\varepsilon_0$, $m_e$ are the plasma density, the electron charge, speed of light, permittivity of free space and the speed of light, respectively. $E_{acc}$ is the acceleration field for the electron beam. To avoid phase slippage over acceleration length $L$, $\delta$ must be less than $\pi$, otherwise the electrons will overrun the protons. For the LHC beam driven plasma wakefield acceleration, the calculation shows that the phase slippage length (or maximum acceleration length) is about 2.1 km (assuming the plasma density of $10^{15}$ cm$^{-3}$ and final proton beam energy is around 0.5 TeV). Therefore for the above $e^+$-$e^-$ collider design, 2 km long acceleration channel meets the phase slippage requirement. For the SPS beam driven plasma wakfield acceleration, we consider two cases, one is to accelerate the electron beam up to 500 GeV and the other to 100 GeV. The phase slippage for the above two cases are shown in Fig. 2. For a 500 GeV electron acceleration case, the final energy of proton beam should be larger than 330 GeV so as to satisfy the phase slippage requirement. If we use the average accelerating (decelerating) field of ~ 1 GeV/m (the plasma density is $10^{15}$ cm$^{-3}$), the maximum dephasing length is about 170 m. This provides the basic parameter to design such an acceleration stage. For a 100 GeV electron beam production, the phase slippage is always in the safe region. Therefore for a SPS drive beam, producing a 100 GeV beam seems reasonable.

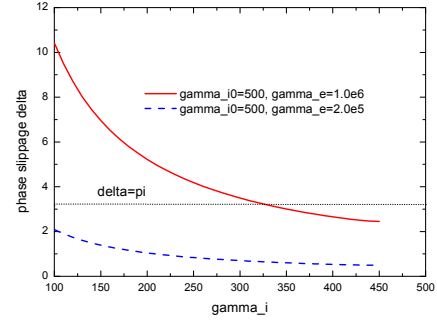

Figure 2: Phase slippage for an SPS proton beam as a function of the final $\gamma_i$ of the proton drive beam for a single stage 500 GeV and 100 GeV electron beam production, respectively.

## CONCLUSION

Simulation shows that a modulated proton bunch can be used to drive a large amplitude plasma wakefield and accelerate the electron beam to high energy. We therefore conceive of an $e^+$-$e^-$ collider and an $e$-$p$ collider design based on this scheme. Using the LHC beam as the drive beam, it is possible to reach a 2 TeV centre-of-mass electron positron collider. For an $e$-$p$ collider design, the SPS beam can be used as the drive beam to accelerate an electron beam up to 100 GeV. The centre-of-mass energy in this case is larger than that of the current LHeC design. Phase slippage between the proton beam and electron (positron) beam may become a limiting factor for the future high energy linear collider design based on the modulated proton-driven plasma wakefield acceleration scheme.